# Low-Rank and Framelet Based Sparsity Decomposition for Interventional MRI Reconstruction

Zhao He, Ya-Nan Zhu, Suhao Qiu, Xiaoqun Zhang, Yuan Feng

*Abstract*— *Objective:* Interventional MRI (i-MRI) is crucial for MR image-guided therapy. Current image reconstruction methods for dynamic MR imaging are mostly retrospective that may not be suitable for i-MRI in real-time. Therefore, an algorithm to reconstruct images without a temporal pattern as in dynamic imaging is needed for i-MRI. *Methods:* We proposed a low-rank and sparsity (LS) decomposition algorithm with framelet transform to reconstruct the interventional feature with a high temporal resolution. Different from the existing LS based algorithm, we utilized the spatial sparsity of both the low-rank and sparsity components. We also used a primal dual fixed point (PDFP) method for optimization of the objective function to avoid solving sub-problems. Intervention experiments with gelatin and brain phantoms were carried out for validation. *Results:* The LS decomposition with framelet transform and PDFP could provide the best reconstruction performance compared with those without. Satisfying reconstruction results were obtained with only 10 radial spokes for a temporal resolution of 60 ms. *Conclusion and Significance:* The proposed method has the potential for i-MRI in many different application scenarios.

*Index Terms*—Interventional MRI, low-rank and sparsity decomposition, image reconstruction, framelet.

## I. INTRODUCTION

INTERVENTIONAL MRI (i-MRI) is a crucial method for MR image-guided therapy. Compared with other imaging modalities such as CT and ultrasound, MR images provide great soft-tissue contrast. This can greatly improve the surgical outcome and patient care in intervention procedures such as biopsy [1] and deep brain stimulation (DBS) [2]-[4]. To monitor the interventional process and track the position of the interventional feature in real time, fast data acquisition and image reconstruction are necessary for i-MRI [5], [6].

Algorithms for fast MR imaging have been evolved from constrained reconstruction [7], generalized-series (GS) method [8]-[10] and keyhole method [11], to a variety of k-t based methods such as UNFOLD [12]. By explaining it in the x-f space, the k-t based methods have been extended to k-t blast [13] and k-t sense [14]. With the compressed sensing (CS) technique, a more general CS based k-t FOCUSS was proposed, using both k-t blast and GS for dynamic reconstruction [15]. Combined with non-Cartesian sampling scheme, CS based k-t method has been successfully used for a variety of dynamic image reconstruction [16], [17]. However, most of the fast MR imaging methods are aimed at retrospective reconstruction of dynamic images. This may not satisfy the high temporal resolution and real-time reconstruction requirements for i-MRI. With the rapid development of deep learning (DL) in recent years, many DL-based methods have been proposed for fast MR imaging [18]-[23]. However, the limited training data available restricted the application of the DL-based methods in i-MRI.

Reconstruction of undersampled data using partially separable (PS) function for low-rank constraint, has been successfully applied in dynamic MRI [24]-[28]. PS techniques have also been used with CS in the reconstruction of quantitative MR images [29], [30]. Using the nonconvex Schatten p-norm for low-rank constraint, Lingala et al. (2011) explored the sparsity and low-rank (SLR) structure in the Karhunen Louve Transform (KLT) domain [31]. With the temporal Fourier transform for the sparsity component, Tremoulheac et al. (2014) proposed the k-t Robust Principle Components (RPCA) to decompose the low-rank and sparsity components with the alternating direction method of multipliers (ADMM) [32]. Otazo et al. (2015) proposed a low-rank and sparsity decomposition method (L+S) using iterative soft thresholding for optimization [33]. Similar decomposition methods were also used in dynamic parallel MR imaging [34] and temperature imaging [35]. Recently, Wang et al. (2020) proposed to include the spatiotemporal regularizer for the sparsity component and adopted a primal-dual algorithm for optimization [36]. Although patch-based reconstruction using spatial information has been proposed [37], [38], most of the algorithms did not consider the spatial compressibility of the low-rank component [39]. In addition, the retrospective reconstruction for dynamic images may not apply for i-MRI, especially in real-time scenarios.

This work was supported by grants 31870941,117711288 from National Natural Science Foundation of China (NSFC), grant 19441907700 from Science and Technology Commission of Shanghai Municipality (STCSM). (Corresponding authors: Yuan Feng and Xiaoqun Zhang.)

Z. He, S. Qiu and Y. Feng are with School of Biomedical Engineering, Shanghai Jiao Tong University, Shanghai, 200030, China (e-mail: fengyuan@sjtu.edu.cn).

Y. Zhu and X. Zhang are with the School of Mathematical Sciences, MOE-LSC and Institute of Natural Sciences, Shanghai Jiao Tong University, Shanghai 200040, China (e-mail: xqzhang@sjtu.edu.cn).

Zhao He and Ya-Nan Zhu contributed equally.

In this study, we proposed a low-rank and sparsity decomposition (LS) with framelet transform (LSF) for reconstruction. The spatial sparsity constraints of the low-rank and sparsity components were included in the LS decomposition. The LSF model was solved with a primal dual fixed point (PDFP) algorithm. Moreover, a group-based reconstruction scheme combined with a golden-angle radial sampling scheme was adopted for the real-time requirement in i-MRI. Simulation and phantom interventional experiments were implemented for validation. Reconstruction results were compared with state-of-the-art methods.

## II. THEORY

*A. Problem formulation*

*1) Low-rank and sparsity decomposition for i-MRI*

The acquired k-space data at time point t is
$$\mathbf{d}(\mathbf{k}, t) = \int I(\mathbf{r}, t) e^{-j2\pi(\mathbf{k}\cdot\mathbf{r})} d\mathbf{r}, \quad (1)$$
where $I(\mathbf{r}, t)$ is the image to be reconstructed, $\mathbf{k}$ is the k-space trajectory, and $\mathbf{r}$ is the spatial location. Treated as a discrete array, $I(\mathbf{r}, t)$ can be written as a Casorati matrix $\mathbf{C} \in \mathbb{C}^{N \times M}$ [25], [27], [30]:
$$\mathbf{C} = \begin{bmatrix} I(\mathbf{r}_1, t_1) & \cdots & I(\mathbf{r}_1, t_M) \\ \vdots & \ddots & \vdots \\ I(\mathbf{r}_N, t_1) & \cdots & I(\mathbf{r}_N, t_M) \end{bmatrix}, \quad (2)$$
where $N$ is the spatial location and $M$ is the temporal point. Therefore, (1) can be written as
$$\mathbf{d} = \mathbf{\Omega F S C} + \mathbf{\varepsilon}, \quad (3)$$
where $\mathbf{S} \in \mathbb{C}^{N \times N}$ is the sensitivity mapping, $\mathbf{F} \in \mathbb{C}^{N \times N}$ is the Fourier encoding, $\mathbf{\Omega} \in \mathbb{C}^{N \times N}$ is the sampling scheme, and $\mathbf{\varepsilon} \in \mathbb{C}^{N \times M}$ is the noise. Reconstruction of $I(\mathbf{r}, t)$ is equivalent of solving $\mathbf{C}$ from the ill-posed problem of (3). To solve $\mathbf{C}$, the low rank constraint can be implemented by either partially separable function decomposition [25], [27], [30], nonconvex Schatten p-norm [31], or nuclear norm [32], [33], [36], [40]. If we use a linear encoding matrix $\mathbf{E} = \mathbf{\Omega F S}$, $\mathbf{E} \in \mathbb{C}^{N \times N}$, the general reconstruction is
$$\hat{\mathbf{C}} = \arg\min_{\mathbf{C}} \|\mathbf{EC} - \mathbf{d}\| + \lambda R(\mathbf{C}), \quad (4)$$
where $R(\cdot)$ is the function to implement low-rank and $\lambda$ is the regularization parameter.

For partially separable function decomposition, the low-rank constraint is implemented by a matrix factorization
$$\mathbf{C} = \mathbf{UV}, \quad (5)$$
where $\mathbf{U} \in \mathbb{C}^{N \times L}$ is the spatial coefficient matrix and $\mathbf{V} \in \mathbb{C}^{L \times M}$ is the temporal basis. For regularization with nuclear norm, the low-rank constraint can be combined with a sparsity decomposition
$$\mathbf{C} = \mathbf{L} + \mathbf{S}, \quad (6)$$
where $\mathbf{L} \in \mathbb{C}^{N \times M}$ is the low-rank matrix and $\mathbf{S} \in \mathbb{C}^{N \times M}$ is the sparse matrix.

*2) The proposed method*

The partially separable (PS) function decomposition has been widely used in dynamic image reconstruction. But the temporal basis could be difficult to decompose for images from i-MRI since there exists no motion pattern. On the other hand, the background and the interventional feature could be separated into a low-rank matrix $\mathbf{L}$ and sparse matrix $\mathbf{S}$ by the LS decomposition. This low-rank plus sparsity model has been used to reconstruct dynamic image, where optimization is carried out with respect to both $\mathbf{L}$ and $\mathbf{S}$ terms [32], [33]:
$$\{\mathbf{L}, \mathbf{S}\} = \arg\min_{\mathbf{L},\mathbf{S}} \tfrac{1}{2} \|\mathbf{E}(\mathbf{L}+\mathbf{S}) - \mathbf{d}\|_2^2 + \lambda_L \|\mathbf{L}\|_* + \lambda_S \|\nabla_t \mathbf{S}\|_1, \quad (7)$$
where $\|\mathbf{L}\|_*$ is the nuclear norm of $\mathbf{L}$, $\nabla_t$ represents a total-variation along the temporal direction of $\mathbf{S}$, $\lambda_L$ and $\lambda_s$ are regularization parameters.

According to the CS theory, both $\mathbf{L}$ and $\mathbf{S}$ were compressible in the spatial domain after the appropriate sparsifying transform such as total variation (TV), wavelets, or framelets. In this work, we adopted the framelet transform that can preserve important image features and provide enough regularization in smooth regions [34], [41]. To utilize the spatial sparsity constraints of the $\mathbf{L}$ and $\mathbf{S}$ components, we proposed a LS decomposition with framelet (LSF) model for i-MRI reconstruction:
$$\{\mathbf{L}, \mathbf{S}\} = \arg\min_{\mathbf{L},\mathbf{S}} \tfrac{1}{2} \|\mathbf{E}(\mathbf{L}+\mathbf{S}) - \mathbf{d}\|_2^2 + \lambda_L \|\mathbf{L}\|_* + \lambda_S \|\nabla_t \mathbf{S}\|_1 + \lambda_L^\psi \|\psi \mathbf{L}\|_1 + \lambda_S^\psi \|\psi \mathbf{S}\|_1. \quad (8)$$
where $\psi$ is the framelet transform, $\lambda_L^\psi$ and $\lambda_S^\psi$ are the regularization parameters. By adding these extra terms, we impose the spatial regularity on both low rank and sparsity components.

*B. Optimization for Reconstruction*

Previous studies have used a combination of singular value thresholding and iterative soft thresholding to optimize (7) [33]. This is equivalent to the proximal gradient method [42]. However, the optimization method involves solving subproblems with TV-clipping, and there are no general rules on how to terminate the subproblem [43]. In this study, we proposed to solve the optimization problem using a primal dual fixed point (PDFP) method. It is easy to implement and can solve (7) and (8) without introducing subproblems.

*1) PDFP algorithm*

Consider the following optimization problem
$$\arg\min_\mathbf{x} f(\mathbf{x}) + g(\mathbf{B}(\mathbf{x})), \quad (9)$$
where $f(\mathbf{x})$ is proper convex l.s.c., and has $1/\beta$-Lipschitz continuous gradient. $g(\mathbf{x})$ is proper convex l.s.c. and may not be differentiable (e.g., $g(\mathbf{x})$ can be $l_1$ norm $\|\cdot\|_1$ or nuclear norm $\|\cdot\|_*$). $\mathbf{B}$ is a linear transform.

Problem (9) is one of the most well-known models in machine learning and imaging science and can be solved by many first-order methods, for example, alternating direction of multipliers (ADMM) [44], [45], primal dual hybrid gradient (PDHG) [46], and primal dual fixed point method (PDFP) [47]. In this study, we use PDFP to avoid solving subproblems. For optimization of problem (9) using PDFP, the procedure is summarized in Algorithm I.

In Algorithm I, $\lambda_{max}(\cdot)$ denotes the maximum eigenvalue of a given matrix. For a convex function $h(\mathbf{x})$, the operator $\text{Prox}_h(\cdot): \mathbb{C}^n \to \mathbb{C}^n$ is defined as

$$\text{Prox}_h(\mathbf{y}) = \arg\min_{\mathbf{x}} h(\mathbf{x}) + \frac{1}{2}\|\mathbf{x}-\mathbf{y}\|_2^2. \quad (10)$$

For example, if $h(\mathbf{x}) = c\|\mathbf{x}\|_1$, $\mathbf{x} \in \mathbb{C}^n$, then for $\mathbf{y} \in \mathbb{C}^n$, $\text{Prox}_h(\mathbf{y})$ is computed by soft thresholding:

$$\begin{aligned}\text{Prox}_h(\mathbf{y}) &= \arg\min_{\mathbf{x}} c\|\mathbf{x}\|_1 + \frac{1}{2}\|\mathbf{x}-\mathbf{y}\|_2^2\\ &= soft(\mathbf{y},c)\\ &= \text{sign}(\mathbf{y}).*\max\{|\mathbf{y}|-c, 0\}.\end{aligned} \quad (11)$$

Let $h^*(\mathbf{y})$ be the conjugate function of $h$ at $\mathbf{y}$ defined by

$$h^*(\mathbf{y}) = \sup_{x\in D}<\mathbf{x},\mathbf{y}> - h(\mathbf{x}). \quad (12)$$

If $h(\mathbf{x}) = c\|\mathbf{x}\|_1$, the conjugate function of $h$ at $\mathbf{y}$ is

$$h^*(\mathbf{y}) = I_C(\mathbf{y}) \quad (13)$$

where $C = \{\mathbf{x}|\|\mathbf{x}\|_\infty \leq c\}$ and $I_C(\cdot)$ is indicator of set C which is defined as

$$I_C(\mathbf{x}) = \begin{cases}0, & \mathbf{x}\in C,\\ +\infty, & \mathbf{x}\notin C.\end{cases} \quad (14)$$

Then,

$$\begin{aligned}\text{Prox}_{h^*}(\mathbf{y}) &= \arg\min_x I_C(\mathbf{x}) + \frac{1}{2}\|\mathbf{x}-\mathbf{y}\|_2^2\\ &= \text{Project}_C(\mathbf{y})\\ &= \mathbf{y}./\max\{|\mathbf{y}/c|, 1\},\end{aligned} \quad (15)$$

where $C = \{\mathbf{x}|\|\mathbf{x}\|_\infty \leq c\}$.

Similarly, if $h(\mathbf{X}) = c\|\mathbf{X}\|_*$, $\mathbf{X} \in \mathbb{C}^{N\times M}$, then for $\mathbf{Y} \in \mathbb{C}^{N\times M}$, $\text{Prox}_h(\mathbf{Y})$ is defined by

$$\text{Prox}_h(\mathbf{Y}) = \mathbf{U}\,\text{diag}(soft(\text{diag}(\mathbf{\Sigma}),c))\mathbf{V}^*, \quad (16)$$

where $\mathbf{X} = \mathbf{U\Sigma V}^*$ is the singular value decomposition of $\mathbf{X}$. And $\text{Prox}_{h^*}(\mathbf{Y})$ is

$$\text{Prox}_{h^*}(\mathbf{Y}) = \mathbf{U}\,\text{diag}(\text{Project}_C(\text{diag}(\mathbf{\Sigma})))\mathbf{V}^* \quad (17)$$

where $C = \{\mathbf{x}|\|\mathbf{x}\|_\infty \leq c\}$.

---

**Algorithm I** PDFP Algorithm

---
**Initialization:** $\mathbf{x}^0, \mathbf{y}^0, \mathbf{p}^0$, $0 < \lambda \leq 1/\lambda_{max}(\mathbf{BB}^T)$ and $0 < \gamma \leq 2\beta$.
**for** $k = 1, 2, \ldots$
1. $\mathbf{y}^{k+1} = \mathbf{x}^k - \gamma\nabla f(\mathbf{x}^k) - \gamma\mathbf{B}^T\mathbf{p}^k$
2. $\mathbf{p}^{k+1} = \text{Prox}_{\frac{\lambda}{\gamma}g^*}\left(\frac{\lambda}{\gamma}\mathbf{By}^{k+1} + \mathbf{p}^k\right)$
3. $\mathbf{x}^{k+1} = \mathbf{x}^k - \gamma\nabla f(\mathbf{x}^k) - \gamma\mathbf{B}^T\mathbf{p}^{k+1}$
**Iterate until the stopping criterion is reached.**

---

*2) LS decomposition with PDFP (LSP)*

The LS model (7) can be formulated as follows:

$$\min_{\mathbf{L,S}} \frac{1}{2}\left\|(\mathbf{E}\ \mathbf{E})\begin{bmatrix}\mathbf{L}\\\mathbf{S}\end{bmatrix} - \mathbf{d}\right\|_2^2 + g\left(\mathbf{B}\begin{pmatrix}\mathbf{L}\\\mathbf{S}\end{pmatrix}\right). \quad (18)$$

where

$$g(x_1,x_2) = f_1(x_1) + f_2(x_2) = \lambda_L\|x_1\|_* + \lambda_S\|x_2\|_1, \quad (19)$$

and

$$\mathbf{B} = \begin{bmatrix}I & 0\\ 0 & \nabla_t\end{bmatrix}. \quad (20)$$

The steps of optimization of (18) using PDFP are summarized in Algorithm II (see Appendix A for details).

In Algorithm II,

$$\text{Prox}_{\frac{\lambda}{\gamma}f_1^*}(\mathbf{A}) = \mathbf{U}\text{diag}(\text{Project}_{C_L}(\text{diag}(\mathbf{S})))\mathbf{V}^*, \quad (21)$$

where $\mathbf{A} = \mathbf{USV}^*$ is the singular value decomposition of $\mathbf{A}$, $C_L = \{x|\|x\|_\infty \leq \lambda_L\}$ and $\text{Project}_{C_L}$ are defined in (11) (replace $c$ by $\lambda_L$).

$$\text{Prox}_{\frac{\lambda}{\gamma}f_2^*}(x) = \text{Project}_{C_S}(x), \quad (22)$$

where $C_S = \{x|\|x\|_\infty \leq \lambda_S\}$.

---

**Algorithm II** LSP model

---
**Initialization:** $\mathbf{y}_L^0, \mathbf{y}_S^0, \mathbf{L}^0, \mathbf{S}^0 \in \mathbb{C}^{N\times M}$, $\mathbf{P}_L^0, \mathbf{P}_S^0 \in \mathbb{C}^{N\times M}$, $0 < \lambda \leq 1/\lambda_{max}(\mathbf{BB}^T)$ and $0 < \gamma \leq 2\beta$.
**for** $k = 1, 2, \ldots$
1. $\mathbf{y}_L^{k+1} = \mathbf{L}^k - \gamma\mathbf{E}^*(\mathbf{E}(\mathbf{L}^k + \mathbf{S}^k) - \mathbf{d}) - \gamma\mathbf{P}_L^k$
2. $\mathbf{y}_S^{k+1} = \mathbf{S}^k - \gamma\mathbf{E}^*(\mathbf{E}(\mathbf{L}^k + \mathbf{S}^k) - \mathbf{d}) - \gamma\nabla_t^T\mathbf{P}_S^k$
3. $\mathbf{P}_L^{k+1} = \text{Prox}_{\frac{\lambda}{\gamma}f_1^*}\left(\frac{\lambda}{\gamma}\mathbf{y}_L^{k+1} + \mathbf{P}_L^k\right)$
4. $\mathbf{P}_S^{k+1} = \text{Prox}_{\frac{\lambda}{\gamma}f_2^*}\left(\frac{\lambda}{\gamma}\nabla_t\mathbf{y}_S^{k+1} + \mathbf{P}_S^k\right)$
5. $\mathbf{L}^{k+1} = \mathbf{L}^k - \gamma\mathbf{E}^*(\mathbf{E}(\mathbf{L}^k + \mathbf{S}^k) - \mathbf{d}) - \gamma\mathbf{P}_L^{k+1}$
6. $\mathbf{S}^{k+1} = \mathbf{S}^k - \gamma\mathbf{E}^*(\mathbf{E}(\mathbf{L}^k + \mathbf{S}^k) - \mathbf{d}) - \gamma\nabla_t^T\mathbf{P}_S^{k+1}$
**Iterate until the stopping criterion is reached.**

---

*3) LSF with PDFP (LSFP)*

Equation (8) can be formulated as:

$$\min_{\mathbf{L,S}} \frac{1}{2}\left\|(\mathbf{E}\ \mathbf{E})\begin{bmatrix}\mathbf{L}\\\mathbf{S}\end{bmatrix} - \mathbf{d}\right\|_2^2 + g\left(\mathbf{B}\begin{pmatrix}\mathbf{L}\\\mathbf{S}\end{pmatrix}\right), \quad (23)$$

where

$$\begin{aligned}g(x_1,x_2,x_3,x_4) &= f_3(x_1) + f_4(x_2) + f_5(x_3) + f_6(x_4)\\ &= \lambda_L\|x_1\|_* + \lambda_S\|x_2\|_1 + \lambda_L^\psi\|x_3\|_1 + \lambda_S^\psi\|x_4\|_1,\end{aligned} \quad (24)$$

and

$$\mathbf{B} = \begin{bmatrix}I & 0\\ 0 & \nabla_t\\ \psi & 0\\ 0 & \psi\end{bmatrix}. \quad (25)$$

The steps to solve LSF model with PDFP are summarized in Algorithm III (see Appendix B for details).

In Algorithm III, $m_1$ is the dimension of range of $\psi$. The proximal operator $\text{Prox}_{\frac{\lambda}{\gamma}f_3^*} \to \text{Prox}_{\frac{\lambda}{\gamma}f_6^*}$ can be derived as in (21) and (22).

---

**Algorithm III** LSFP model

---
**Initialization:** $\mathbf{y}_L^0, \mathbf{y}_S^0, \mathbf{L}^0, \mathbf{S}^0 \in \mathbb{R}^{N\times M}$, $\mathbf{P}_L^0, \mathbf{P}_S^0 \in \mathbb{R}^{N\times M}$, $\mathbf{P}_{\psi_L}^0, \mathbf{P}_{\psi_S}^0 \in \mathbb{R}^{N_1\times M}$, $0 < \lambda \leq 1/\lambda_{max}(\mathbf{BB}^T)$ and $0 < \gamma \leq 2\beta$.
**for** $k = 1, 2, \ldots$
1. $\mathbf{y}_L^{k+1} = \mathbf{L}^k - \gamma\mathbf{E}^*(\mathbf{E}(\mathbf{L}^k + \mathbf{S}^k) - \mathbf{d}) - \gamma\mathbf{P}_L^k - \gamma\psi^T\mathbf{P}_{\psi_L}^k$
2. $\mathbf{y}_S^{k+1} = \mathbf{S}^k - \gamma\mathbf{E}^*(\mathbf{E}(\mathbf{L}^k + \mathbf{S}^k) - \mathbf{d}) - \gamma\nabla_t^T\mathbf{P}_S^k - \gamma\psi^T\mathbf{P}_{\psi_S}^k$
3. $\mathbf{P}_L^{k+1} = \text{Prox}_{\frac{\lambda}{\gamma}f_3^*}\left(\frac{\lambda}{\gamma}\mathbf{y}_L^{k+1} + \mathbf{P}_L^k\right)$
4. $\mathbf{P}_S^{k+1} = \text{Prox}_{\frac{\lambda}{\gamma}f_4^*}\left(\frac{\lambda}{\gamma}\nabla_t\mathbf{y}_S^{k+1} + \mathbf{P}_S^k\right)$
5. $\mathbf{P}_{\psi_L}^{k+1} = \text{Prox}_{\frac{\lambda}{\gamma}f_5^*}\left(\frac{\lambda}{\gamma}\psi\mathbf{y}_L^{k+1} + \mathbf{P}_{\psi_L}^k\right)$
6. $\mathbf{P}_{\psi_S}^{k+1} = \text{Prox}_{\frac{\lambda}{\gamma}f_6^*}\left(\frac{\lambda}{\gamma}\psi\mathbf{y}_S^{k+1} + \mathbf{P}_{\psi_S}^k\right)$
7. $\mathbf{L}^{k+1} = \mathbf{L}^k - \gamma\mathbf{E}^*(\mathbf{E}(\mathbf{L}^k + \mathbf{S}^k) - \mathbf{d}) - \gamma\mathbf{P}_L^{k+1} - \gamma\psi^T\mathbf{P}_{\psi_L}^{k+1}$
8. $\mathbf{S}^{k+1} = \mathbf{S}^k - \gamma\mathbf{E}^*(\mathbf{E}(\mathbf{L}^k + \mathbf{S}^k) - \mathbf{d}) - \gamma\nabla_t^T\mathbf{P}_S^{k+1} - \gamma\psi^T\mathbf{P}_{\psi_S}^{k+1}$
**Iterate until the stopping criterion is reached.**

## III. METHODS

### A. Data acquisition and reconstruction methods

To achieve fast i-MRI, we adopted a golden-angle radial sampling scheme (Fig. 1(a)). K-space data was continuously acquired during the intervention process. For dynamic imaging, the reconstruction is carried out after collecting all k-space data throughout the dynamic process (Fig. 1(b)). Each frame is reconstructed based on a certain number of radial spokes. This retrospective reconstruction is not suitable for i-MRI, especially with a real-time requirement. In this study, we proposed a group-based scheme to reconstruct images from i-MRI (Fig. 1(c)). Rather than a retrospective reconstruction using all acquired k-space data, we divided the continuously acquired radial spokes into $M$ groups, each of them can reconstruct $n$ images. If only 10 radial spokes are used for reconstructing one frame, and we take 5 frames for each reconstruction group, only 50 radial spokes are needed for a reconstruction of one group. This highly undersampled scheme with an acceleration factor of 40 provides a potential way for real-time i-MRI.

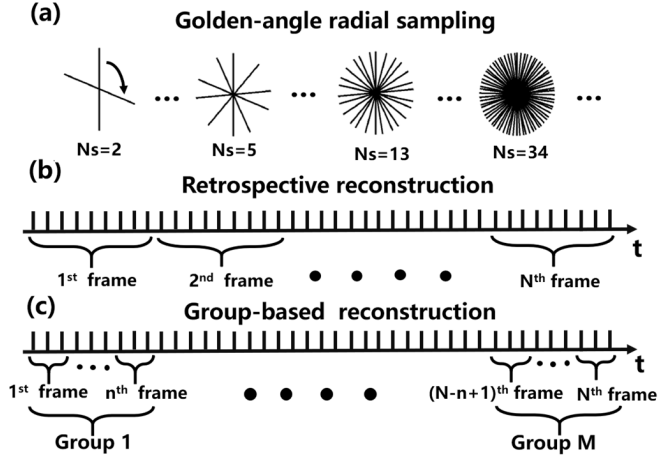

Fig. 1. Illustrations of the data acquisition and reconstruction scheme. (a) A continuous golden-angle radial sampling method used for i-MRI in this study (golden angle = 111.25°). (b) Conventional dynamic image reconstruction based on a retrospective scheme. (c) The proposed group-based reconstruction method for real-time i-MRI reconstruction.

### B. A simulation of brain intervention

To evaluate the proposed reconstruction method, 200 brain intervention images were generated from a reference T2-weighted MR image. The images had a matrix size of 200×200 with 11 channels (Fig. 2). The nonuniform fast Fourier transform (NUFFT) was applied to simulate the radial sampling. The k-space data was acquired with 400 readout points, a total of 2000 radial spokes, and 11 channels.

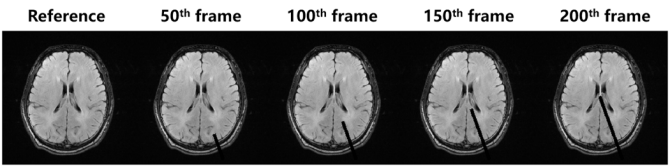

Fig. 2. A simulation of brain intervention showing the reference image and 4 of the 200 simulated images.

### C. Interventional experiments

To further evaluate the proposed algorithm, interventional experiments were carried out with the proposed imaging method. We used a homogeneous gelatin phantom and a porcine brain for the i-MRI experiment. The gelatin phantom made of a mixture of gelatin and glycerol [48] was put in a cylinder container with a diameter of 120 mm (Fig. 3(a)).

To simulate the brain intervention, the porcine brain was fixed in a gelatin phantom embedded and the whole phantom was placed in a 3D-printed human skull (Fig. 3(b)). A custom-built interventional device with a needle was used for the intervention (Fig. 3(c)). The gelatin phantom and the head model were placed inside the head coil for interventional imaging (Fig. 3(d)). All the imaging experiments were carried out on a 3T MRI scanner (uMR 790, United Imaging Healthcare, Shanghai, China).

Before the intervention, T1-weighted (T1W), T2-weighted (T2W), and fully sampled reference images were acquired. The reference images were used to estimate the 17 coil sensitivity maps. The scanning parameters were summarized in TABLE I. For both phantoms, a total of 2000 golden-angle radial spokes were acquired during the intervention, which took a total of 9.2 s for the gelatin phantom and 12 s for the brain phantom.

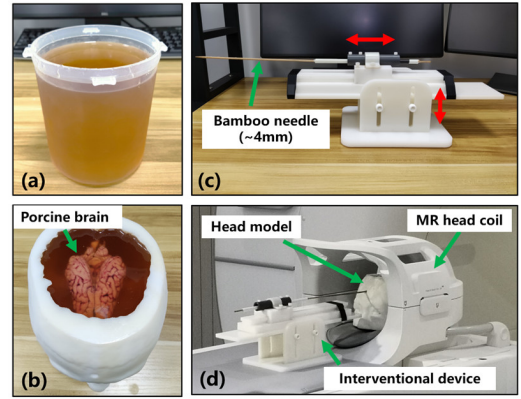

Fig. 3. Phantoms and devices for i-MRI experiments. (a) A homogeneous gel phantom for intervention. (b) A simulated brain phantom for intervention. A porcine brain was embedded in a gelatin phantom in a 3D printed human skull model. (c) A custom-built 2-DOF interventional device with a bamboo needle. The diameter of the needle was ~4mm. (d) The experiment setup of the i-MRI.

TABLE I
SCANNING PARAMETERS FOR BOTH GELATIN AND BRAIN PHANTOMS

| | T1W-FSE | T2W-FSE | Reference | Intervention |
|---|---|---|---|---|
| Sampling points in the readout (2×) | 512 | 512 | 512 | 512 |
| Sampling radial spokes | ~ | ~ | 402 | 2000 |
| Slice thickness (mm) | 5 | 5 | 5 | 5 |
| FOV (mm²) | 300×300 | 300×300 | 300×300 | 300×300 |
| TR/TE (ms) | 2000/10.22 | 8000/102.6 | 4.6/1.9, 6/2.94 | 4.6/1.9, 6/2.94 |
| Flip angle (degree) | 135 | 150 | 30 | 30 |
| Bandwidth (Hz/pixel) | 180 | 300 | 300 | 300 |
| Acquisition time (s) | 62 | 56 | 1.8, 2.4 | 9.2, 12 |

## D. Comparison and evaluation of algorithms

To evaluate the performance of the proposed method, the reconstruction results from LSFP are compared with those from algorithms such as Golden-angle RAdial Sparse Parallel MRI (GRASP) [16], LS [33], and LSP. The GRASP algorithm solves the following optimization problem:

$$\hat{\mathbf{C}} = \arg\min_{\mathbf{C}} \|\mathbf{E}\mathbf{C} - \mathbf{d}\|_2^2 + \lambda\|\nabla_t \mathbf{C}\|_1. \quad (26)$$

Additionally, to verify the advantage of using framelet transform in LSFP, TV (LSTP) and wavelet (LSWP) transforms were also used for comparison. Finally, to investigate the influences of the spatial sparsity of the **L** and **S** in model (8), the following two models optimized with the PDFP algorithm were used for comparisons:

$$\{\mathbf{L}, \mathbf{S}\} = \arg\min_{\mathbf{L},\mathbf{S}} \frac{1}{2}\|\mathbf{E}(\mathbf{L} + \mathbf{S}) - \mathbf{d}\|_2^2 + \lambda_L\|\mathbf{L}\|_* + \lambda_s\|\nabla_t \mathbf{S}\|_1 + \lambda_L^{\psi}\|\psi\mathbf{L}\|_1. \quad (27)$$

$$\{\mathbf{L}, \mathbf{S}\} = \arg\min_{\mathbf{L},\mathbf{S}} \frac{1}{2}\|\mathbf{E}(\mathbf{L} + \mathbf{S}) - \mathbf{d}\|_2^2 + \lambda_L\|\mathbf{L}\|_* + \lambda_s\|\nabla_t \mathbf{S}\|_1 + \lambda_s^{\psi}\|\psi\mathbf{S}\|_1. \quad (28)$$

To evaluate the reconstruction performance of each algorithm, we calculated the normalized mean square error (NMSE), peak signal-to-noise ratio (PSNR), and structure similarity (SSIM) for the reconstruction of simulated and phantom images. All images and error maps were scaled to [0, 1]. The definitions of these metrics can be found in Appendix C.

The algorithms were implemented with MATLAB R2020a on an Ubuntu 20.04 LTS (64-bit) operating system equipped with an AMD Ryzen 9 5950x central processing unit (CPU) and NVIDIA RTX 3090 graphics processing unit (GPU, 24 GB memory).

## IV. RESULTS

### A. Simulated intervention

Reconstruction of the simulated interventional images shows that the LSFP had the least streak artifacts and noise than the state-of-the-art methods (Fig. 4). Quantitatively, the proposed LSFP approach yielded the lowest NMSE, the highest PSNR, and SSIM (TABLE II).

### B. Gelatin phantom i-MRI

To evaluate the performance of the proposed method, we used images reconstructed retrospectively with the LS algorithm as ground truth (Fig. 5). The evaluation metrics of the reconstruction were summarized in TABLE III. The metric values of the reconstructed images were similar at different time points. This demonstrates the robustness of the proposed method.

We also compared the reconstruction results with different combinations of spokes per frame (SPF) and frames per group (FPG) (Fig. 6). When FPG was 5 and 10, and SPF was 5, 8, and 10, the reconstruction performance improved with the increase of FPG or SPF (TABLE IV). This is because as FPG or SPF increased, more data was used for reconstruction. However, the reconstruction time increased with SFP and FPG.

A set of interventional MR images of the gelatin phantom were reconstructed by NUFFT, GRASP, LS, LSP, and LSFP methods (Fig. 7). With an acceleration factor around 40, the proposed LSFP yielded the least artifacts and noise. The NMSE, PSNR, and SSIM for LSFP method were 0.0041, 28.2509, and 0.7521, respectively (TABLE V).

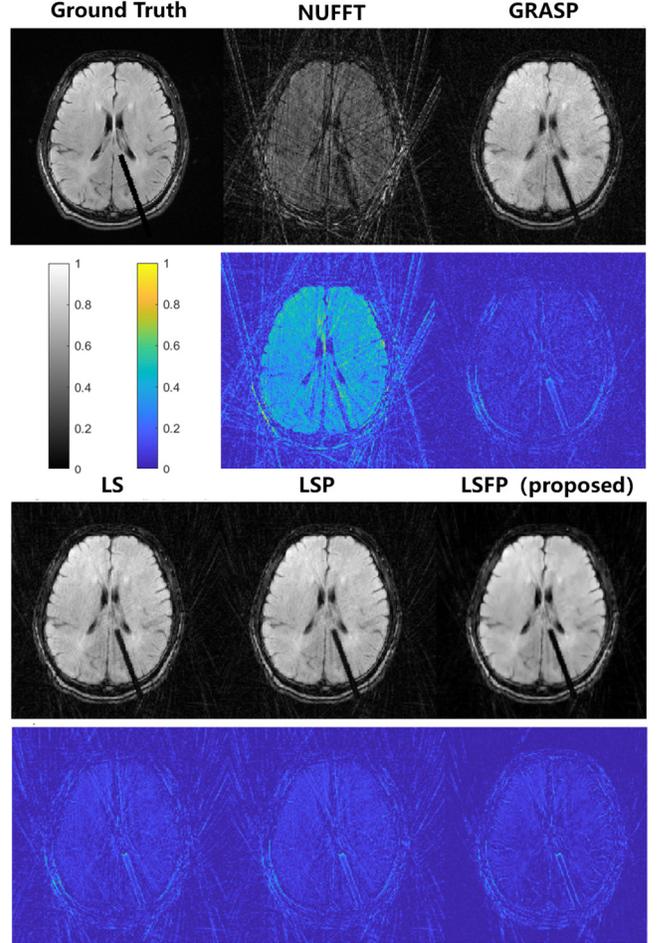

Fig. 4. A comparison of different algorithms. The ground truth is the 150th simulated brain intervention image. A group-based reconstruction strategy (SPF=10, FPG=5, a total of 200 frames for 2000 spokes) was adopted for reconstruction using NUFFT, GRASP, LS, LSP, and LSFP algorithms. The acceleration factor is about 40. Movies of the interventional images can be found in Supporting Information Video S1.

TABLE II
A COMPARISON OF EVALUATION METRICS USING DIFFERENT ALGORITHMS
(SFP=10, FPG=5)

|      | NUFFT  | GRASP  | LS     | LSP    | LSFP   |
|------|--------|--------|--------|--------|--------|
| NMSE | 0.3189 | 0.0299 | 0.0284 | 0.0270 | **0.0177** |
| PSNR | 13.4785 | 23.7571 | 23.9855 | 24.2019 | **26.0239** |
| SSIM | 0.2322 | 0.5784 | 0.6308 | 0.6497 | **0.7523** |

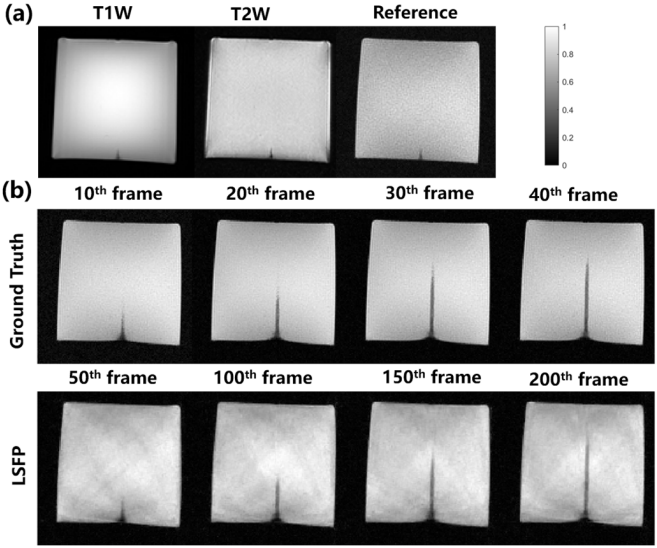

Fig. 5. MR images of the gel phantom in the interventional experiment. (a)T1W, T2W, and fully sampled reference images before the intervention. The first row of (b) are the retrospectively reconstructed ground truth images with the LS method (SPF=50, a total 40 frames for 2000 spokes). The second row of (b) are reconstructed images with the LSFP algorithm (SFP=10, FPG=5, a total of 200 frames for 2000 spokes).

TABLE III
QUANTITATIVE EVALUATION OF DIFFERENT FRAMES RECONSTRUCTED WITH THE LSFP ALGORITHM (SFP=10, FPG=5)

|  | 50th frame | 100th frame | 150th frame | 200th frame | Mean ± STD |
|---|---|---|---|---|---|
| NMSE | 0.0059 | 0.0059 | 0.0041 | 0.0050 | 0.0052 ± 0.0009 |
| PSNR | 26.4061 | 26.6382 | 28.2509 | 27.3709 | 27.1665 ± 0.8317 |
| SSIM | 0.6881 | 0.7397 | 0.7521 | 0.7301 | 0.7275 ± 0.0278 |

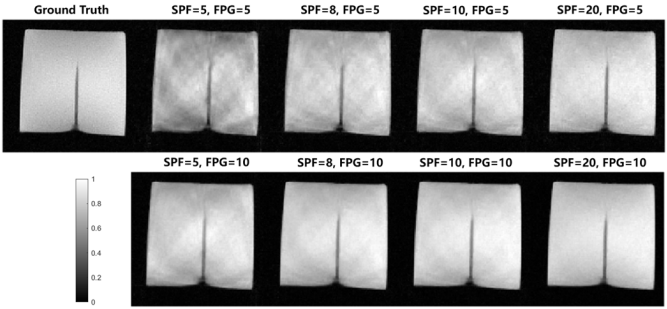

Fig. 6. A comparison of reconstruction results from different SPF and FPG values using the proposed algorithm. The 40th frame of the ground truth was selected for this illustration.

TABLE IV
EVALUATION METRICS USING DIFFERENT COMBINATIONS OF SPF AND FPG

| **FPG=5** | SPF=5 | SPF=8 | SPF=10 | SPF=20 |
|---|---|---|---|---|
| NMSE | 0.0266 | 0.0066 | 0.0050 | 0.0052 |
| PSNR | 20.1407 | 26.2194 | 27.3709 | 27.2150 |
| SSIM | 0.6442 | 0.7085 | 0.7301 | 0.7859 |
| **FPG=10** | SPF=5 | SPF=8 | SPF=10 | SPF=20 |
| NMSE | 0.0057 | 0.0049 | 0.0071 | 0.0046 |
| PSNR | 26.8455 | 27.4782 | 25.8660 | 27.7219 |
| SSIM | 0.7457 | 0.7884 | 0.8039 | 0.8433 |

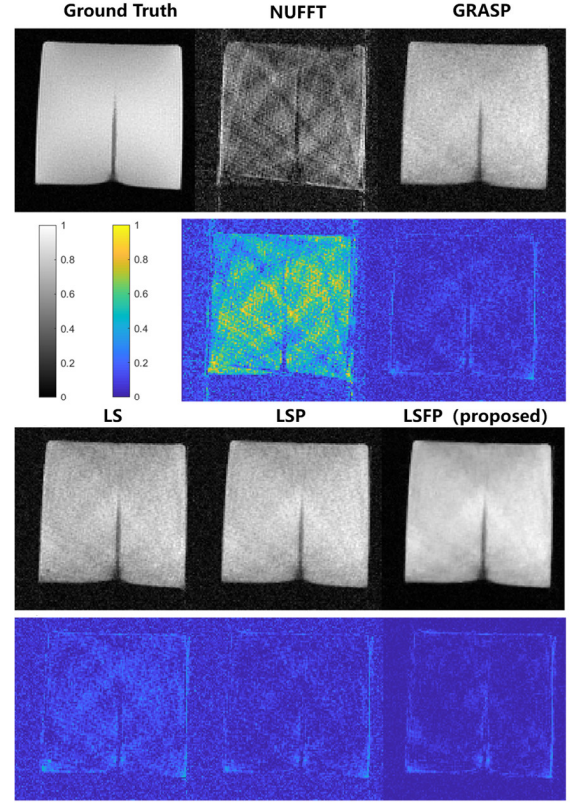

Fig. 7. A comparison of reconstructed interventional images using NUFFT, GRASP, LS, LSP, and LSFP algorithms. The ground truth was reconstructed with the LS algorithm retrospectively (SPF=50, a total of 40 frames for 2000 spokes). NUFFT, GRASP, LS, LSP, and LSFP algorithms used group-based reconstruction strategy (SPF=10, FPG=5, a total of 200 frames for 2000 spokes). The 30th frame of the ground truth was selected for this illustration. Movies of the interventional images can be found in Supporting Information Video S2.

TABLE V
A COMPARISON OF EVALUATION METRICS USING DIFFERENT ALGORITHMS (SFP=10, FPG=5)

|  | NUFFT | GRASP | LS | LSP | LSFP |
|---|---|---|---|---|---|
| NMSE | 0.2714 | 0.0138 | 0.0205 | 0.0112 | **0.0041** |
| PSNR | 10.0006 | 22.9495 | 21.2248 | 23.8507 | **28.2509** |
| SSIM | 0.1696 | 0.5007 | 0.4913 | 0.5211 | **0.7521** |

### C. Brain phantom i-MRI

Using LSFP, we observed a clear trajectory of the needle during the intervention (Fig. 8). The evaluation metrics of the proposed method were summarized in TABLE VI. The metric values of the reconstructed images were similar at different time points.

Different combinations SPF and FPG were compared qualitatively (Fig. 9) and quantitatively (TABLE VII). The reconstruction performance improved with an increase of FPG when SPF was fixed. Different from the gelatin phantom experiment, with the same spokes for one group, the reconstruction performance increased with increased SPF values. This suggests that a careful selection of parameters is needed to obtain the best reconstruction quality.

Compared with other methods, LSFP generated the least artifacts in the i-MRI of the brain phantom (Fig. 10). In terms of NMSE, PSNR, and SSIM, LSFP also had the best performance (TABLE VIII). Notably, LSFP overperformed LSTP and LSWP. This indicates that compared with TV and wavelet transform, the framelet transform could provide improved reconstruction results.

The reconstruction based on models (27) and (28) showed the NMSE, PSNR, and SSIM were 0.0051, 31.2491, 0.6743, and 0.0042, 32.0502, 0.8201, respectively (Fig. S1). These performance metrics were between those of LSTP and LSWP but superior to that without spatial sparsity regularizations (TABLE VIII). It showed that the spatial sparsity of both the low-rank and sparse components, together with the choice of framelet transform were important for the overall reconstruction performance.

TABLE VII
A COMPARISON OF EVALUATION METRICS USING DIFFERENT COMBINATIONS OF SPF AND FPG

| **FPG=5** | SPF=5 | SPF=8 | SPF=10 | SPF=20 |
|---|---|---|---|---|
| NMSE | 0.0108 | 0.0047 | 0.0041 | 0.0027 |
| PSNR | 27.9870 | 31.5795 | 32.2306 | 33.9666 |
| SSIM | 0.8443 | 0.8759 | 0.8890 | 0.8980 |
| **FPG=10** | SPF=5 | SPF=8 | SPF=10 | SPF=20 |
| NMSE | 0.0049 | 0.0035 | 0.0032 | 0.0021 |
| PSNR | 31.4572 | 32.8672 | 33.2479 | 35.1325 |
| SSIM | 0.8584 | 0.8521 | 0.8589 | 0.8446 |

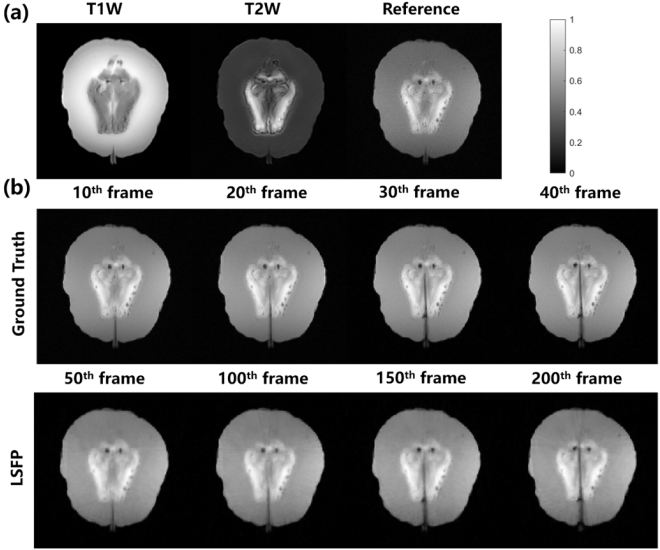

Fig. 8. MR images of the brain phantom in the interventional experiment. (a) T1W, T2W, and fully sampled reference images before the intervention. (b) Retrospectively reconstructed images as ground truth (SFP=50, a total of 40 frames for 2000 spokes), and the images reconstructed with the LSFP (SPF=10, FPG=5, total 200 frames for 2000 spokes).

TABLE VI
EVALUATION METRIC VALUES OF IMAGES RECONSTRUCTED WITH LSFP (SFP=10, FPG=5)

|  | 50th frame | 100th frame | 150th frame | 200th frame | Mean ± STD |
|---|---|---|---|---|---|
| NMSE | 0.0033 | 0.0031 | 0.0026 | 0.0041 | 0.0033 ± 0.0006 |
| PSNR | 33.3387 | 33.5059 | 34.1957 | 32.2306 | 33.3177 ± 0.8142 |
| SSIM | 0.8003 | 0.8353 | 0.9049 | 0.8890 | 0.8573 ± 0.0483 |

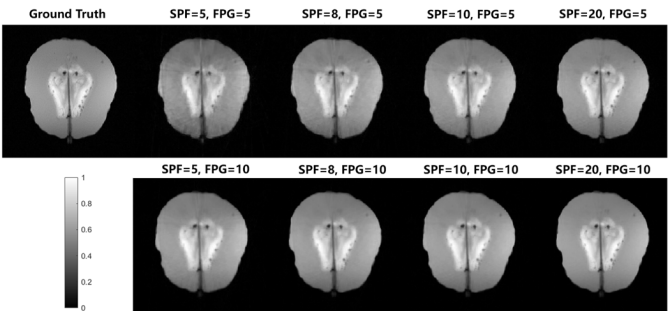

Fig. 9. A comparison of reconstruction results from different SPF and FPG values with the proposed algorithm. The 40th frame of the ground truth was selected for this illustration.

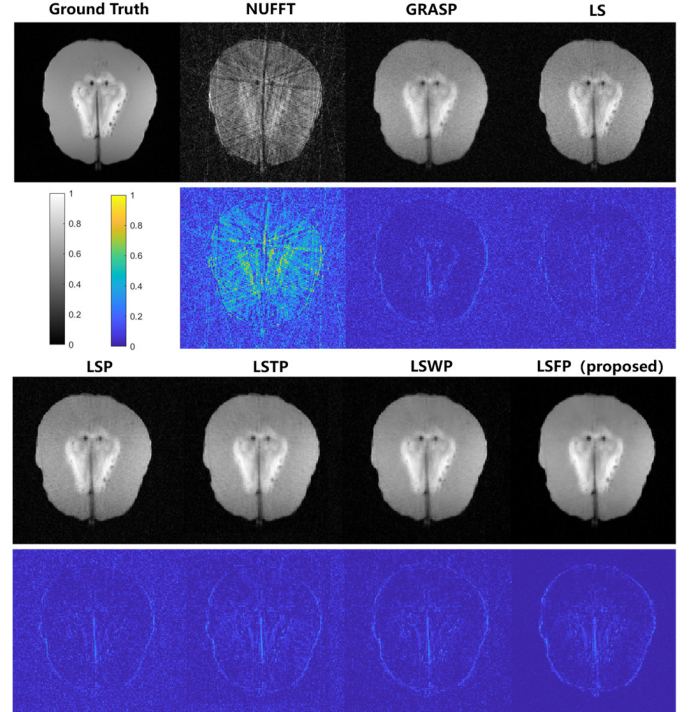

Fig. 10. A comparison of different algorithms. The ground truth was reconstructed with the LS decomposition retrospectively (SPF=50, a total of 40 frames for 2000 spokes). A group-based reconstruction strategy was also adopted for NUFFT, GRASP, LS, LSP, LSTP, LSWP and LSFP algorithms (SPF=10, FPG=5, a total of 200 frames for 2000 spokes). The 30th frame of the ground truth was selected for this illustration. Movies of the interventional images can be found in Supporting Information Video S3.

TABLE VIII
A COMPARISON OF EVALUATION METRICS USING DIFFERENT ALGORITHMS (SFP=10, FPG=5)

|  | NUFFT | GRASP | LS | LSP | LSTP | LSWP | LSFP |
|---|---|---|---|---|---|---|---|
| NMSE | 0.1486 | 0.0101 | 0.0091 | 0.0071 | 0.0066 | 0.0044 | **0.0026** |
| PSNR | 16.5831 | 28.2771 | 28.7008 | 29.7774 | 30.1298 | 31.8806 | **34.1957** |
| SSIM | 0.1795 | 0.5387 | 0.5442 | 0.6118 | 0.6769 | 0.7554 | **0.9049** |

V. DISCUSSION

For most of the LS decomposition algorithms, only temporal sparsification was used. Our study is different from that Wang et al. (2020) who employed a total generalized variation (TGV) as the spatiotemporal regularizer for the sparse component [36], and Qu et al. (2014) who proposed the patch-based reconstruction method to exploit local redundancies and low-rank matrix structures [37], [38]. In this study, simultaneous spatial sparsity constraints for the low-rank and sparse components in LS decomposition algorithms were explored. Therefore, using the framelet transform for spatial sparsification of both low rank and sparsity components, we proposed the LSFP model. The framelet are known to be effective in recovering higher-order singularities from limited data. Theoretically, total variation regularization can be viewed as a special case of framelet by setting Haar types of filters [41], [49]. By choosing piecewise linear or cubic filters, more complex structures could be captured, leading to improved image quality. Besides the superiority of framelet transforms that was shown numerically for medical image restoration tasks [50], here we also showed that the adoption of framelet could lead to better reconstruction results for i-MRI. In addition, many optimization algorithms for the LS decomposition involved solving subproblems [32], [33]. To avoid solving the complex subproblems that lack a general termination criterium, we proposed to solve the LS decomposition with a primal dual fixed point (PDFP) algorithm rather than the commonly used ADMM or SVT.

We have found that the proposed method is not very sensitive to the selection of regularization parameters. In this work, regularization parameters $\lambda_L$, $\lambda_S$, $\lambda_L^\psi$, and $\lambda_S^\psi$ were carefully tuned to acquire the best reconstruction performance. Based on the parameter selection procedure of Otazo et al. (2015), parameters $\lambda_L$ and $\lambda_S$ were first selected. Then, $\lambda_L^\psi$, and $\lambda_S^\psi$ were empirically selected for the simulation data. For gelatin phantom and porcine brain phantom intervention data, we only needed to fine-tune the parameters based on the tuned parameters from the simulation data.

In interventional surgery, the subject motion and deformation may introduce potential artifacts to the reconstruction. To evaluate the motion and deformation of the phantom after the intervention, the differences between the first and last interventional images of the two interventional experiments were calculated (Fig. 11). We observed that the potential motion of the background occurred mostly around the needle, but the background was not affected. This indicated that the deformation mostly resulted from the needle insertion around the needle region.

Different from the retrospective reconstruction for dynamic MRI [29]-[33], [36], we adopted a group-based reconstruction scheme with only 50 radial spokes for fast imaging. Therefore, for real-time imaging, the practical temporal resolution was ~0.3s. Although the interventional feature can be well reconstructed with 5 spokes, the results showed that 10 spokes generated the better performance. In terms of the reconstructed image frames, the theoretical temporal resolution was ~0.06s for 10 radial spokes.

Although many data-based machine learning algorithms were proposed for the reconstruction of MR images [18]-[23], [51], it may be challenging to solve the generalization problem. For example, a trained model in brain cases may not be applicable in the abdomen region. Most importantly, the training data could be hard to obtain for i-MRI cases, since the intervention is an invasive procedure. Therefore, analytical methods that avoid all the above concerns could provide a robust and general way for image reconstruction in i-MRI.

This study has some limitations. Although group-based reconstruction could greatly increase the temporal resolution, the reconstruction is still "retrospective" within one group. The proposed LSFP method requires longer machine time due to additional regularization terms using framelet transform. However, we note that many strategies can be adopted to reduce the overall reconstruction time. For example, the usage of gpuNUFFT toolbox based on the graphics processing units (GPU) [52] can achieve ~4 times acceleration factor compared with the classical NUFFT toolbox [53] (TABLE IX). Furthermore, using the accelerated PDFP method [54] may also accelerate reconstruction, which will be included in our future work.

In addition, we only investigated a few grouping parameters. There may exist an optimized grouping strategy for different application scenarios. Future studies include implementation of the algorithm in GPU and using parallel computing strategies, and application in different interventional cases.

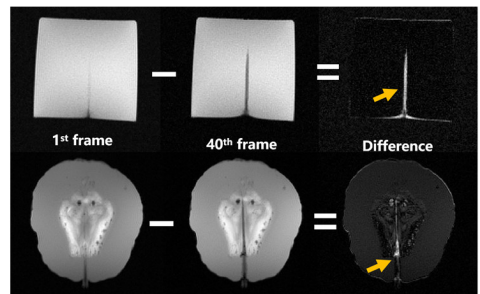

Fig. 11. An evaluation of the background deformation. The image differences were calculated between the 1st frame and the 40th frame for both gelatin and brain i-MRI.

TABLE IX
RECONSTRUCTION TIME FOR ONE GROUP OF PORCINE BRAIN PHANTOM (SFP=10, FPG=5, MATRIX SIZE OF 512 × 512, 17 CHANNELS, 10 ITERATIONS)

| Time (s) | GRASP | LS | LSP | LSTP | LSWP | LSFP |
|---|---|---|---|---|---|---|
| NUFFT [53] | 43.77 | 22.03 | 28.89 | 30.71 | 47.96 | 38.62 |
| gpuNUFFT [52] | 6.82 | 2.19 | 3.14 | 5.38 | 20.77 | 10.14 |
| Acceleration | 6.42 | 10.06 | 9.2 | 5.71 | 2.30 | 3.81 |

## VI. Conclusion

In this study, we proposed a new method for accelerating i-MRI by combining the improved LS decomposition algorithm with golden-angle radial sampling and group-based reconstruction methods. The proposed method takes advantage of the spatial sparsity of the low-rank and sparsity components using framelet transform for fast i-MRI reconstruction. To avoid subproblems, a PDFP algorithm was used for optimization. Gelatin and brain phantom experiments showed the robustness of the proposed method. Evaluation metrics also showed the LSFP algorithm had better performance than other retrospective reconstruction methods. The improved temporal resolution demonstrates the potential of the proposed method for real-time i-MRI.

## APPENDIX A

### DERIVATION OF LSP ALGORITHM

Denote

$$\mathbf{x}^k = \begin{bmatrix} \mathbf{L}^k \\ \mathbf{S}^k \end{bmatrix}, \mathbf{y}^k = \begin{bmatrix} \mathbf{y}_\mathbf{L}^k \\ \mathbf{y}_\mathbf{S}^k \end{bmatrix}, \mathbf{p}^k = \begin{bmatrix} \mathbf{P}_\mathbf{L}^k \\ \mathbf{P}_\mathbf{S}^k \end{bmatrix}, g^* = f_1^* + f_2^*, \quad (29)$$

and $\mathbf{B}$ is defined in (20).

The PDFP for (18) is updated in the following three steps:

1) $\mathbf{y}^{k+1} = \mathbf{x}^k - \gamma \nabla f(\mathbf{x}^k) - \gamma \mathbf{B}^T \mathbf{p}^k$ is equivalent to

$$\begin{bmatrix} \mathbf{y}_\mathbf{L}^{k+1} \\ \mathbf{y}_\mathbf{S}^{k+1} \end{bmatrix} = \begin{bmatrix} \mathbf{L}^k \\ \mathbf{S}^k \end{bmatrix} - \gamma \begin{bmatrix} \mathbf{E}^*(\mathbf{E}(\mathbf{L}^k + \mathbf{S}^k) - \mathbf{d}) \\ \mathbf{E}^*(\mathbf{E}(\mathbf{L}^k + \mathbf{S}^k) - \mathbf{d}) \end{bmatrix} - \gamma \mathbf{B}^T \begin{bmatrix} \mathbf{P}_\mathbf{L}^k \\ \mathbf{P}_\mathbf{S}^k \end{bmatrix}$$
$$= \begin{bmatrix} \mathbf{L}^k \\ \mathbf{S}^k \end{bmatrix} - \gamma \begin{bmatrix} \mathbf{E}^*(\mathbf{E}(\mathbf{L}^k + \mathbf{S}^k) - \mathbf{d}) \\ \mathbf{E}^*(\mathbf{E}(\mathbf{L}^k + \mathbf{S}^k) - \mathbf{d}) \end{bmatrix} - \gamma \begin{bmatrix} \mathbf{P}_\mathbf{L}^k \\ \nabla_t^T \mathbf{P}_\mathbf{S}^k \end{bmatrix}. \quad (30)$$

2) To compute $\mathbf{p}^{k+1} = \text{Prox}_{\frac{\lambda}{\gamma}g^*}\left(\frac{\lambda}{\gamma}\mathbf{B}\mathbf{y}^{k+1} + \mathbf{p}^k\right)$, note that

$$\frac{\lambda}{\gamma}\mathbf{B}\mathbf{y}^{k+1} + \mathbf{p}^k = \frac{\lambda}{\gamma}\mathbf{B}\begin{bmatrix} \mathbf{y}_\mathbf{L}^{k+1} \\ \mathbf{y}_\mathbf{S}^{k+1} \end{bmatrix} + \begin{bmatrix} \mathbf{P}_\mathbf{L}^k \\ \mathbf{P}_\mathbf{S}^k \end{bmatrix}$$
$$= \begin{bmatrix} \frac{\lambda}{\gamma}\mathbf{y}_\mathbf{L}^{k+1} + \mathbf{P}_\mathbf{L}^k \\ \frac{\lambda}{\gamma}\nabla_t \mathbf{y}_\mathbf{S}^{k+1} + \mathbf{P}_\mathbf{S}^k \end{bmatrix}, \quad (31)$$

Then

$$\begin{bmatrix} \mathbf{P}_\mathbf{L}^{k+1} \\ \mathbf{P}_\mathbf{S}^{k+1} \end{bmatrix} = \begin{bmatrix} \text{Prox}_{\frac{\lambda}{\gamma}f_1^*}\left(\frac{\lambda}{\gamma}\mathbf{y}_\mathbf{L}^{k+1} + \mathbf{P}_\mathbf{L}^k\right) \\ \text{Prox}_{\frac{\lambda}{\gamma}f_2^*}\left(\frac{\lambda}{\gamma}\nabla_t \mathbf{y}_\mathbf{S}^{k+1} + \mathbf{P}_\mathbf{S}^k\right) \end{bmatrix}. \quad (32)$$

3) Similar to 1), $\mathbf{x}^{k+1} = \mathbf{x}^k - \gamma \nabla f(\mathbf{x}^k) - \gamma \mathbf{B}^T \mathbf{p}^{k+1}$ is equivalent to

$$\begin{bmatrix} \mathbf{L}^{k+1} \\ \mathbf{S}^{k+1} \end{bmatrix} = \begin{bmatrix} \mathbf{L}^k \\ \mathbf{S}^k \end{bmatrix} - \gamma \begin{bmatrix} \mathbf{E}^*(\mathbf{E}(\mathbf{L}^k + \mathbf{S}^k) - \mathbf{d}) \\ \mathbf{E}^*(\mathbf{E}(\mathbf{L}^k + \mathbf{S}^k) - \mathbf{d}) \end{bmatrix} - \gamma \begin{bmatrix} \mathbf{P}_\mathbf{L}^{k+1} \\ \nabla_t^T \mathbf{P}_\mathbf{S}^{k+1} \end{bmatrix}. \quad (33)$$

## APPENDIX B

### DERIVATION OF LSFP ALGORITHM

Denote

$$\mathbf{x}^k = \begin{bmatrix} \mathbf{L}^k \\ \mathbf{S}^k \end{bmatrix}, \mathbf{y}^k = \begin{bmatrix} \mathbf{y}_\mathbf{L}^k \\ \mathbf{y}_\mathbf{S}^k \end{bmatrix}, \mathbf{p}^k = \begin{bmatrix} \mathbf{P}_\mathbf{L}^k \\ \mathbf{P}_\mathbf{S}^k \\ \mathbf{P}_{\psi_\mathbf{L}}^k \\ \mathbf{P}_{\psi_\mathbf{S}}^k \end{bmatrix}, \quad (34)$$

$$g^* = f_3^* + f_4^* + f_5^* + f_6^*,$$

and $\mathbf{B}$ is defined in (25).

The PDFP for (23) is updated in the following three steps:

1) $\mathbf{y}^{k+1} = \mathbf{x}^k - \gamma \nabla f(\mathbf{x}^k) - \gamma \mathbf{B}^T \mathbf{p}^k$ is equivalent to

$$\begin{bmatrix} \mathbf{y}_\mathbf{L}^{k+1} \\ \mathbf{y}_\mathbf{S}^{k+1} \end{bmatrix} = \begin{bmatrix} \mathbf{L}^k \\ \mathbf{S}^k \end{bmatrix} - \gamma \begin{bmatrix} \mathbf{E}^*(\mathbf{E}(\mathbf{L}^k + \mathbf{S}^k) - \mathbf{d}) \\ \mathbf{E}^*(\mathbf{E}(\mathbf{L}^k + \mathbf{S}^k) - \mathbf{d}) \end{bmatrix} - \gamma \mathbf{B}^T \begin{bmatrix} \mathbf{P}_\mathbf{L}^k \\ \mathbf{P}_\mathbf{S}^k \\ \mathbf{P}_{\psi_\mathbf{L}}^k \\ \mathbf{P}_{\psi_\mathbf{S}}^k \end{bmatrix}$$
$$= \begin{bmatrix} \mathbf{L}^k \\ \mathbf{S}^k \end{bmatrix} - \gamma \begin{bmatrix} \mathbf{E}^*(\mathbf{E}(\mathbf{L}^k + \mathbf{S}^k) - \mathbf{d}) \\ \mathbf{E}^*(\mathbf{E}(\mathbf{L}^k + \mathbf{S}^k) - \mathbf{d}) \end{bmatrix}$$
$$- \gamma \begin{bmatrix} \mathbf{P}_\mathbf{L}^k + \psi^T \mathbf{P}_{\psi_\mathbf{L}}^k \\ \nabla_t^T \mathbf{P}_\mathbf{S}^k + \psi^T \mathbf{P}_{\psi_\mathbf{S}}^k \end{bmatrix}. \quad (35)$$

2) To compute $\mathbf{p}^{k+1} = \text{Prox}_{\frac{\lambda}{\gamma}g^*}\left(\frac{\lambda}{\gamma}\mathbf{B}\mathbf{y}^{k+1} + \mathbf{p}^k\right)$, note that

$$\frac{\lambda}{\gamma}\mathbf{B}\mathbf{y}^{k+1} + \mathbf{p}^k = \frac{\lambda}{\gamma}\mathbf{B}\begin{bmatrix} \mathbf{y}_\mathbf{L}^{k+1} \\ \mathbf{y}_\mathbf{S}^{k+1} \end{bmatrix} + \begin{bmatrix} \mathbf{P}_\mathbf{L}^k \\ \mathbf{P}_\mathbf{S}^k \\ \mathbf{P}_{\psi_\mathbf{L}}^k \\ \mathbf{P}_{\psi_\mathbf{S}}^k \end{bmatrix}$$
$$= \begin{bmatrix} \frac{\lambda}{\gamma}\mathbf{y}_\mathbf{L}^{k+1} + \mathbf{P}_\mathbf{L}^k \\ \frac{\lambda}{\gamma}\nabla_t \mathbf{y}_\mathbf{S}^{k+1} + \mathbf{P}_\mathbf{S}^k \\ \frac{\lambda}{\gamma}\psi \mathbf{y}_\mathbf{L}^{k+1} + \mathbf{P}_{\psi_\mathbf{L}}^k \\ \frac{\lambda}{\gamma}\psi \mathbf{y}_\mathbf{S}^{k+1} + \mathbf{P}_{\psi_\mathbf{S}}^k \end{bmatrix}, \quad (36)$$

Then

$$\begin{bmatrix} \mathbf{P}_\mathbf{L}^{k+1} \\ \mathbf{P}_\mathbf{S}^{k+1} \\ \mathbf{P}_{\psi_\mathbf{L}}^{k+1} \\ \mathbf{P}_{\psi_\mathbf{S}}^{k+1} \end{bmatrix} = \begin{bmatrix} \text{Prox}_{\frac{\lambda}{\gamma}f_3^*}\left(\frac{\lambda}{\gamma}\mathbf{y}_\mathbf{L}^{k+1} + \mathbf{P}_\mathbf{L}^k\right) \\ \text{Prox}_{\frac{\lambda}{\gamma}f_4^*}\left(\frac{\lambda}{\gamma}\nabla_t \mathbf{y}_\mathbf{S}^{k+1} + \mathbf{P}_\mathbf{S}^k\right) \\ \text{Prox}_{\frac{\lambda}{\gamma}f_5^*}\left(\frac{\lambda}{\gamma}\psi \mathbf{y}_\mathbf{L}^{k+1} + \mathbf{P}_{\psi_\mathbf{L}}^k\right) \\ \text{Prox}_{\frac{\lambda}{\gamma}f_6^*}\left(\frac{\lambda}{\gamma}\psi \mathbf{y}_\mathbf{S}^{k+1} + \mathbf{P}_{\psi_\mathbf{S}}^k\right) \end{bmatrix}. \quad (37)$$

3) Similar to 1), $\mathbf{x}^{k+1} = \mathbf{x}^k - \gamma \nabla f(\mathbf{x}^k) - \gamma \mathbf{B}^T \mathbf{p}^{k+1}$ is equivalent to

$$\begin{bmatrix} \mathbf{L}^{k+1} \\ \mathbf{S}^{k+1} \end{bmatrix} = \begin{bmatrix} \mathbf{L}^k \\ \mathbf{S}^k \end{bmatrix} - \gamma \begin{bmatrix} \mathbf{E}^*(\mathbf{E}(\mathbf{L}^k + \mathbf{S}^k) - \mathbf{d}) \\ \mathbf{E}^*(\mathbf{E}(\mathbf{L}^k + \mathbf{S}^k) - \mathbf{d}) \end{bmatrix} -$$
$$\gamma \begin{bmatrix} \mathbf{P}_\mathbf{L}^{k+1} + \psi^T \mathbf{P}_{\psi_\mathbf{L}}^{k+1} \\ \nabla_t^T \mathbf{P}_\mathbf{S}^{k+1} + \psi^T \mathbf{P}_{\psi_\mathbf{S}}^{k+1} \end{bmatrix}. \quad (38)$$

## APPENDIX C

### DEFINITIONS OF EVALUATION METRICS

The normalized mean square error (NMSE), peak signal-to-noise ratio (PSNR), and structure similarity (SSIM) are defined as follows.

NMSE is defined as

$$\text{NMSE} = \frac{\sum_{j=1}^{J}\sum_{k=1}^{K}[\mathbf{X}(j,k) - \hat{\mathbf{X}}(j,k)]^2}{\sum_{j=1}^{J}\sum_{k=1}^{K}[\mathbf{X}(j,k)]^2}, \quad (39)$$

where $J$ and $K$ are the row and column numbers of images $\mathbf{X}$ and $\hat{\mathbf{X}}$.

PSNR is defined as

$$\text{PSNR} = 10\log_{10}\left(\frac{\text{MAX}^2}{\text{MSE}}\right), \quad (40)$$

where MAX is the maximum value of $\mathbf{X}$ and

$$\text{MSE} = \frac{\sum_{j=1}^{J}\sum_{k=1}^{K}[\mathbf{X}(j,k) - \hat{\mathbf{X}}(j,k)]^2}{JK}. \quad (41)$$

SSIM is defined as

$$\text{SSIM}(\mathbf{X}, \mathbf{Y}) = \frac{(2u_X u_Y + C_1)(2\sigma_{XY} + C_2)}{(u_X^2 + u_Y^2 + C_1)(\sigma_X^2 + \sigma_Y^2 + C_2)}, \quad (42)$$

where $u_X$ and $u_Y$ are the mean values of images $\mathbf{X}$ and $\mathbf{Y}$, respectively. $\sigma_X$ and $\sigma_Y$ are the standard deviation of $\mathbf{X}$ and $\mathbf{Y}$. $\sigma_{XY}$ is the convariance of $\mathbf{X}$ and $\mathbf{Y}$. $C_1 = (K_1 L)^2$ and $C_2 = (K_2 L)^2$ are constants where $L$ is the dynamic range, which is 255 for 8-bit gray-scale images. $K_1 = 0.01$ and $K_2 = 0.03$ are suggested parameter values [56], [57].

### ACKNOWLEDGMENT

We thank Prof. Zhi-Pei Liang from UIUC for helpful discussions.